\documentclass[12pt]{article}
\begin{document}                
\begin{center}
{\Huge\bf Primordial flares, flux tubes, MHD waves in the early universe and 
genesis of cosmic gamma ray bursts}
\end{center}
\begin{center}
{\Large\bf K. M. Hiremath}
\end{center}
\begin{center}
{\em Indian Institute of Astrophysics, Bangalore-560034, India,
E-mail : hiremath@iiap.ernet.in}
\end{center}
\begin{abstract}                
It is conjectured that energy sources of the gamma ray bursts are similar to
energy sources which trigger solar and stellar transient activity phenomena
like flares, plasma accelerated flows in the flux tubes and, dissipation
of energy and acceleration of particles by the MHD waves. 
Phenomenologically we examine in detail the following energy sources which 
may trigger gamma ray bursts : (i) cosmic primordial flares which 
could be solar flare like phenomena in the region of inter galactic
or inter galactic cluster regions, (ii) primordial magnetic flux tubes 
that might have been formed from the convective collapse of the primordial 
magnetic flux (iii) nonlinear interaction and dissipation of MHD waves 
that are produced from the perturbations of large-scale inter galactic or 
inter cluster magnetic field of primordial origin. We examine in detail each of 
the afore mentioned phenomena keeping in mind that whether such processes are 
responsible for energy sources  of the gamma ray bursts.
\vskip 0.1cm
By considering the similarity of observations and prevailing physical
conditions in the cosmic environment, the following study suggests that 
most likely and a promising energy source for creation of the gamma ray bursts may be 
due to primordial flares. 
\end{abstract}
\section{Introduction}               

 Since the discovery of the Gamma Ray bursts (GRB), physics of these enigmatic
phenomena remain elusive till today. To date, there are more than hundred 
theoretical models on GRB ( {\em} see the bibliography web site {\em http://ssl.berkeley.  edu/ipn3/index.html} compiled by Hurley). 
Based on insufficient data, in the beginning period 
of observations, many models on GRB favored energy source of galactic origin.
However, recent observations are not consistent with these models 
and the models which explain that source of GRB could be of cosmic origin 
are most favorable (Piran 1999).  Though, these models can explain some of the 
observed phenomena, other outstanding observed properties such as energy
source and isotropic distribution of GRB in the sky remain to  be explained. 

 It is interesting to note that most of the observed properties of the GRB
are almost similar to the observed properties of the solar flares. 
Energetics of other phenomena such as transient flows in the sunspot and 
heating of the  solar corona are worthwhile for comparison with
the energetics of GRB phenomena. It is to be noted that most of the these solar
activity phenomena are directly or indirectly involved with the
 the ambient magnetic field. Hence, in the following study and, in an analogy
with the solar transient activity phenomena, it is supposed
that large scale inter galactic or inter galactic cluster magnetic field
may be of primordial origin could be responsible in creating the
observed cosmic GRB.
 
Presently it is believed (Priest 1981;  Haish and Strong 1991 ; Parker 1994) that 
the source of energy produced in the solar flares is due to phenomenon called {\it magnetic 
reconnection} in a very compact region wherein oppositely directed magnetic 
flux, in the limit of finite electric conductivity, annihilate each other
and releasing required amount of flare energy with the acceleration
of highly energetic particles (Lenters and Miller 1998, Tsuneta and
Naito 1998). Steady Evershed flows of $\sim 10^{5} cm sec^{-1}$ in sunspots
is interpreted as follows : (i) for overall equilibrium of current
free core and current sheath of a flux tube, a upward or downward flow
in the current sheath is essential (Gokhale and  Hiremath 1986) and 
(ii) flows along a flux tube acts like siphon which is driven by pressure 
differences at the end of flux tube foot points ( Montesinos and Thomas 1997)
. On the other hand, transient flows in the sunspots are either due to 
changing conditions at the foot points (Thomas 1994) or due to convective 
inter change of flux tubes in the penumbra (Schlichenmaier 1997).
Heating of the solar corona (Sakurai 1996) is based mainly on the notion 
that MHD waves interact nonlinearly producing shocks which not only heat the 
ambient medium but also accelerate the particles (Miller 1997). 

From the similarity between solar transient activity phenomena
and GRB and, inspired by the theoretical works on magnetic reconnection that explain 
these phenomena, we conjecture that GRB activity could be of similar origin 
in the distant universe. In the following study, we investigate in detail
whether any such physical phenomena in the cosmic environment
are responsible for the creation of the GRB. It is found from
this investigation that, primordial flares (in the regions of
inter galactic or inter galactic clusters), {\it i.e.,} solar flare like 
phenomenon, may be most promising energy source for the creation of GRB. 
Similar studies (Qu Qin and Wang 1977; Vahia and Rao 1988) indicate
relevance of stellar flares as the energy source of GRB. However,
 observed spectra of GRB implies that 
GRB must be of cosmic origin and hence it is unlikely that stellar flares 
may be energy source for the cosmic GRB.

In section 2, we briefly describe the theory of magnetic reconnection.
Application of magnetic reconnection for the explanation of GRB is presented 
in section 2.1. In section 3, we present plasma flows from the primordial
flux tubes. We present physics of the MHD waves and their 
dissipation in the GRB environment in section 4. Results and conclusions are 
presented in the last section. Preliminary results of this study have been
presented at the IAU symposium 195 (Hiremath 1999) and details are
presented in the following paper. 

\section{  Theory of magnetic reconnection }
Let oppositely directed magnetic flux of large length scale $L$ 
are merged with  inflow velocity $v_{in}$. This merging of flux will form a current
sheath of thickness $\delta$. Then the law of magnetic induction according
to MHD description is 
$$
{{\partial B } \over{\partial t}} = curl ( v \times B) + \eta {\nabla}^2 B \, ,  \eqno (1)
$$
where $B$ is strength of the magnetic field, $v$ is the velocity and $\eta$ 
is the magnetic diffusivity of the plasma. The first term in the above equation
is due to convective flow and second term represents the magnetic diffusion
of the plasma. Outside the region of reconnection, magnetic diffusivity
is very low and magnetic field is glued to the plasma and moves
along with the plasma. That is magnetic field is frozen to
the plasma. This condition of infinite electric conductivity fails in
the region of magnetic field reconnection by producing very high
gradients of currents and electric fields. Dissipation of these
strong currents leads to annihilation of magnetic field 
in the region of magnetic reconnection where  
 steady state exists so that convective and resistive
terms in equation (1) are equal. The amount of energy released by 
the annihilation of magnetic field $B$ and cube of length $L$
is estimated to be $\sim$ $L^{3}B^{2}$. The ratio of convective to
resistive term called "magnetic Reynolds number" is given as follows
$$
R_{m} = {v_{in} \delta \over { \eta}}       \, , \eqno (2)  
$$
where $\delta$ is the thickness of reconnection region. The assumption of 
incompressibility and conservation of flux yields
$$
v_{out} = {L \over {\delta}} v_{in} = v_a    \, , \eqno (3)
$$
where $v_{out}$ is out flow velocity, $v_{in}$ is the inflow
velocity  and,  $v_a$ is the ambient Alfven 
wave  velocity whose perturbations are
perpendicular to the field line and travel along it. 
\bigskip
Outside the region of magnetic reconnection, the convective term
in equation (1) dominates over the resistive term. That is $R_{m}>>1$ and
hence the electric field is non-dissipative. However, inside the layer of 
thickness $\delta$, $R_{m}<<1$ and
, the electric field $E = \eta J$ is dissipative which leads to 
domination of kinetic effects.  For example, two dimensional simulations 
(Brown 1999 and references there in) indicate that magnetic flux and 
electron flow decouple in the accelerating region resulting in 
acceleration of electron beams  by ejecting with super-Alfvenic velocity.
\subsection{  Primordial flares}
When I say primordial flares, it means that flares of cosmic
origin in the optically thin medium situated either in the regions of inter galactic
or inter galactic clusters. The length scales are of cosmic 
dimension and, cosmic flares are produced by the oppositely directed magnetic 
flux due to peculiar motion of  cosmic bodies such as galaxies or cluster of 
galaxies or near the sites of supernova ejections from the galaxy.
The region of magnetic reconnection is assumed to be formed
by merging of oppositely directed magnetic field which is
of primordial origin. The typical observed peculiar velocity fields
of galaxies are $\sim 10^7$ cm/sec and length scales of magnetic elements $\sim 10^{24} $cms.
It is expected that in order to reproduce the observed magnetic fields
 $\sim \mu$ G in galaxies, galaxy clusters and inter galactic
clusters (Kroneberg 1994; Vallee {\it et. al} 1987; Lesch and Chiba 1997; Bagchi {\it et. al} 1998;
Taylor and Perely 1993), a primordial field of $\sim 10^{-9} $ G is required.
\smallskip

 Based on the simple flare model described
in section 2 and available observed informations on GRB, I 
estimate the approximate size of the region involved  and
the required amount of energy released in GRB. I also estimate
the strength of the magnetic field and thickness of  the reconnection layer
at the site of GRB formation.  Using this information and conservation
of mass and flux, I estimate strength of the magnetic field of region 
before the formation of the GRB. Then I compare estimated strength
of magnetic field and velocity of outflow $v_{out}$ from the site
of flare region with the strength of the 
observed large-scale magnetic field and peculiar velocity of the galaxies.
\smallskip

We know that the amount of energy released by annihilation of magnetic
field of strength $B$ and cube of length scale $L$ is $\sim L^3 B^2$. 
Observations (Meegan {\it et.al.} 1997) show that total amount of energy 
released by GRB activity is $\sim 10^{53} \, to \, 10^{56}$ ergs if the
GRB sources are at cosmic distance. In order to produce maximum energy of
$\sim 10^{56}$, the length $L$ of the cube should be $\sim 10^{12}$ cms
 and field strength in the reconnection region to be $\sim 10^{10} $ G. In fact, 
observations (Piran 1999) of GRB spectrum in absorption lines does indicate that field of 
$10^{12}$ G is required for explanation of the GRB spectrum  of non thermal 
origin.
\smallskip

If we assume $ L = \delta $, then from equation (3), this implies 
that $v_{in} = v_{out} = v_{a} $. Alfven velocity $v_{a}$ can be 
calculated from the inferred magnetic field ($\sim 10^{10} G$) and
observed maximum time scale ($\sim 1000 \, sec$) of the GRB. This gives the
Alfven velocity of $\sim 10^9 $ cm/sec and density of the reconnection
region $\sim 10 \, g/cm^{-3}$. Note that $v_{in} = v_{a}$ is 100 times more than
the observed velocity ($\sim 10^7 $ cm/sec) of the galaxies.
This calculation is also in contradiction with the conventional
flare mechanisms (Petschek 1964 ) wherein we have $v_{out} = v_{a} = 0.1 v_{in}$.
This indicates that thickness of the reconnecting layer must
be less than length scale $L$.  
If we accept that $v_{out} = 0.1 v_{in}$ then law of conservation
of mass yields density in the vicinity of the in flowing region to be 
$\sim 1 \, gm cm^{-3}$.
\smallskip

Let us now reverse back, using inferred length scale and strength of 
magnetic field in region of reconnection, and ask what could be the strength of
initial magnetic field that might have been distributed over
large length scales. If so, does it match with strength of the observed 
large-scale magnetic fields in the universe. Conservation of
flux leads to a relation 
$$
B_{initial} = B_{present} {L^{2} \over R_{initial}^{2}}   \, , \eqno (4) 
$$
where $B_{initial}$ is the initial strength of the magnetic
field, $B_{present}$ is strength of magnetic field in the
reconnection region and $R_{initial}$ is initial length scale
of the magnetic field. If we take the observed large-scale
magnetic field then for typical length scale $R_{in} \sim 10 \, kpc \sim 10^{22} cm$,
we get $B_{initial} \sim 10^{-10}$ G which is almost similar
to the expected field strength of primordial origin from the observations.
Hence the typical rate of GRB is expected to be $\sim R_{initial}/v_{in}
 \sim 10^{5}$ years which is very close to the observed
rate of GRB $\sim 10^6$ years. 
\section{  Primordial flux tubes }
Let us  assume that magnetic flux tubes, similar to solar flux tubes, 
were formed by the convective collapse of the magnetic flux
in the early history of the universe when the matter was
dominated over the radiation. If those flux tubes  of primordial
origin survive upto the present age, then length scale of such a
flux tube is given as follows
$$
L = (\tau \eta)^{1/2} \, , \eqno (5)
$$
where $\tau$ is survival life time of the flux tube and $\eta$ is the magnetic
diffusivity. This equation is derived from the equation of magnetic
induction in which it is assumed that the directions of velocity of the flux 
tube and the direction of ambient magnetic fields are parallel
to each other. By taking survival life time equivalent to age 
of the universe ($\sim 10 \, billion \, yrs \sim 10^{17} sec$) and 
turbulent magnetic diffusivity $\sim 10^{32} cm^{2}sec^{-1}$
 ({\it eg.} equation 3.29 from Lesch and Chiba 1997), we get length scale 
$L$ of the primordial flux tube is $\sim 10^{25}$ cms. 
In the calculation of turbulent magnetic diffusivity we take 
plasma velocity to be the observed peculiar velocity of galaxies
($\sim\, 10^{7} cms \, sec^{-1}$) and typical intergalactic length scales
of 1kpc. 
\smallskip

If such an isolated and untwisted flux tube form with a current free 
core bounded by thin current sheath, is in an overall equilibrium
with stratified atmosphere, then there is a dominant flow along
the thin current sheath and a net downward force on the core or
vice-versa (Gokhale and Hiremath, 1986). The total amount of energy 
released in time $t$ of such a plasma
flow in the current sheath of a flux tube is given as follows 
$$
E \sim\, ({B^2 \over 4\pi})({\Delta x})t \, , \eqno(6)
$$
where $B$ is strength of the magnetic field in the flux tube, $\Delta x$ is
the total thickness of the current sheath
In the case of sun, typical sunspots have total area of 100 millionths of
hemisphere and typical size of umbra is $\sim\,20\%$ of
size of the total area. If we take length scale of the solar flux to be
size of the convective envelope which is $\sim\, 10^{10}$ cms, then the 
sunspots radii which consist of umbra and penumbra must be $\sim \, 1\%$
of the length of the flux tube. We copy similar picture and physical
properties of the solar flux tube to cosmic flux tube. Assuming
that current sheath is wholly concentrated in penumbral region, simple
calculations yield radius of cosmic flux tube be $\sim\, 10^{23}$ cms
and total thickness of current sheath which is situated between umbra
and penumbral boundary to be $\sim \, 10^{56}\,cm^{2}$.
After substituting these values in equation (6),
we have $E \sim 10^{56} B^{2} t$. That means in order to get required
energy for GRB phenomena in a time scale of 1 sec, we require
 magnetic field strength $B$ of the flux tube of $\sim 1$ G.
\section{  MHD waves in the cosmic environment }
One more expected source of energy for the production of GRB is the dissipation of 
MHD waves which travel along and perpendicular to the large-scale cosmic
magnetic field of primordial origin. Some of these MHD waves are 
believed to be responsible for heating of the solar corona (Kuperus, {\it, et.al.}
 1981, Sakurai 1996). 
\smallskip

For uniform 
magnetic field and for incompressibility, we have Alfven waves whose
perturbations are perpendicular to the magnetic field and travel
along the magnetic filed lines. First order perturbations show that
amplitudes of perturbations never become nonlinear since there are no 
density or pressure perturbations involved with these waves and hence never
dissipate. They simply escape to the space of inter galactic or inter
cluster medium. However when oppositely traveling Alfven waves meet
each other resulting beat period between two waves (Wentzel 1976 ) convert 
their density
fluctuations into slow mode MHD waves (explained in the following text)
and then dissipate their energy. 
Following Wentzel (1976) we compute in the following energy flux released
by the dissipation and length scales of dissipation of the MHD waves.
The flux F due to dissipation is given as follows
$$
F = {B^2 \over {8\pi}} A {{v_a^2} \over {A^2}}  \, , \eqno (7)
$$
where A is amplitude of the waves and $v_{a}$ is Alfven velocity.
By taking the typical observed peak flux of GRB980425 which is of $\sim 10^{-7}  ergs cm^{-2}sec^{-1}$
, we require the amplitude of waves to be $\sim 0.01 cm sec^{-1}$.
In this calculation, ambient density is taken from the observed
density $\sim 10^{-31} gm\, cm^{-3}$ and strength of uniform magnetic
field is taken to be $\sim 10^{-10}$ G (strength of large scale magnetic
field of primordial origin) and Alfven velocity turns out to be 
$\sim 10^{6}\, cm\, sec^{-1}$. The length scale over which these
Alfven waves are dissipated is $\sim 10^{-18} cms$. Thus it
is natural to expect that in the cosmic environment pure Alfven waves can not
 be the candidates for source of the GRB. 
\smallskip

For uniform magnetic field and compressive plasma, the medium
of magnetic plasma has following three types of waves. The {\it shear
Alfven waves} which are similar to the Alfven waves as discussed above.
\smallskip

The {\it fast MHD waves} in which the thermal and the magnetic compressions
are in phase. And finally, the {\it slow MHD waves} whose thermal
and magnetic perturbations are in out of phase. For very weak magnetic field
, $ v_{s} >> v_{a}$ and $v_{f} = v_{s}$, while for a strong magnetic field
$v_{s} << v_{a}$ and $v_{f} = v_{a}$. Here, $v_{s}$ 
is the sound velocity if there is no magnetic field, $v_{f}$ represents
velocity of the fast MHD wave and $v_{a}$ is the Alfven wave velocity. 
The fast waves
move perpendicular to magnetic field lines. In the cosmic environment,
 assuming that the ambient medium consists of
hydrogen plasma only, we get the ambient sound speed for the ambient
temperature $\sim$ of 10$^{\circ}$K, to be  $\sim 10^{4} cm sec^{-1}$ which
is smaller by factor of 100 compared to the Alfven wave velocity. In this
context, the maximum energy flux that can be dissipated from the fast MHD waves
is $\sim \rho A^{2} v_{a}$, where $A$ is amplitude of the wave. Hence, in order to get GRB flux, the
required amplitude of waves should be $\sim$ $ 10^{9}\, cm\, sec^{-1}$.
The length scale of dissipation of this flux is $\sim 1.7 cms$.
\smallskip

In case of {\it slow MHD waves}, thermal and magnetic perturbations
are out of phase. These waves can travel only in direction close to
the direction of the magnetic field. For very week fields, $v_{s}>>v_{a}$
and for strong fields $v_{s} << v_{a}$, then waves travel along 
the direction of field lines only. In case of intermediate field,
 $v_{slow} = v_{a}$, where $v_{slow}$ is the velocity of slow MHD wave,
 then the allowed direction of travel lies between
a cone with half an angle $\theta = 27^{o} $. The wave energy flux
of the slow modes is $\sim \rho A^{2} v_{slow}$. Then the required 
amplitudes of the waves should be $\sim 10^{10}\, cm\, sec^{-1}$.
The length scales over which theses slow waves dissipate their
energy is $\sim 10^{3} cms$.
\section{ Conclusions and discussion }
Our conclusions of this study are as follows :
\smallskip

(i)  Assuming that energy sources of the gamma ray bursts are similar to
the energy sources which trigger solar and stellar transient activity phenomena
like flares, plasma accelerated flows in the flux tubes and, dissipation
of energy by the MHD waves, we estimated physical parameters in
the cosmic environment and compared with the observations.
\smallskip

(ii) If  primordial flares are responsible for the creation of GRB, the  
 following physical parameters are needed in order to explain the
observed maximum amount of energy released by the GRB. The source
region requires length scale of $\sim 10^{12} cms$, strength   
of the annihilating magnetic field $\sim 10^{10}$ G and the inflow
velocity of the plasma may be $\sim 10^{9} cm/sec^{-1}$ if we
assume that thickness is equal to length scale of the reconnection
region. By the theory of conventional flare mechanism and conservation
of mass, we also estimated density of the reconnection region to be
$\sim 10 \, g cm^{-3}$. Finally, by taking the observed typical
length scale of the cosmic environment and conservation of magnetic  
flux, we estimated the strength of the primordial magnetic field
which is almost similar to expected strength of the magnetic field
of primordial origin. Lastly, we estimated typical rate of GRB
per galaxy to be $\sim 10^{5} $ yrs.
\smallskip

(ii) If plasma flows (that dissipate and release required amount 
of energy ) in the primordial flux tubes are responsible for
the creation of GRB, the required length of the flux tube may
be $\sim 10^{25}$ cms. Assuming that total amount of energy released
from the current sheath of the primordial flow is same as the
amount of energy released in GRB, it is estimated that the
magnetic field strength of the flux tube may be $\sim\, 1$ G.
\smallskip

(iii) Finally, we considered nonlinear interaction and dissipation of MHD waves 
in the cosmic environment and estimated flux of the energy released.
It is expected that, in addition to thermal sound waves, cosmic
environment with magnetic field can accommodate Alfven waves, shear
Alfven waves, fast and slow MHD waves. In case of Alfven waves
and in order to get the observed GRB flux, it is estimated that
the amplitudes of the waves that dissipate into shocks may
be $\sim 0.01\, cm\, sec^{-1}$. Similarly, in case of fast and slow
MHD waves the required amplitudes of the waves may be $\sim 10^{9}\, cm\, sec^{-1}$
and $\sim 10^{10}\, cm\, sec^{-1}$.

\bigskip
Most of the phenomena explained in this study appear to be
promising sources of energy for the creation of GRB. However,
depending upon the observational constraints, one can accept
or reject these phenomena.
 For example, among the MHD waves,
if we assume that Alfven waves are responsible for the GRB
phenomena, then estimated dissipation lengths over which
shock be formed is very small. For example, observations
of GRB bursts temporal variability on a time scale $\Delta t \sim 10 \, msec$
implies that source size  should be less than 3000 kms (Piran 1999).
Though, dissipation length is within the observational limit, it is very 
difficult to believe that such low amplitude Alfven waves create
GRB.   
\smallskip

In case of either fast or slow modes though both the waves form
the shock, dissipate the required amount of GRB flux, magnitude
of the amplitude of the waves involved in dissipation is enormous
and author is not aware as to whether observations do indicate
any such amplitudes of either thermal or magnetic perturbations.
Since, density perturbations are involved in fast and slow
MHD waves, it will be interesting if such signatures are detected
in future observations.
\smallskip

Primordial flux tubes are also good candidates for the acceleration
of matter to the required energy of GRB. The difficulty in this
case is how to dissipate the accelerated plasma matter and moreover stability
of such flux tubes (in cosmic environment) should be examined
carefully. Future observations of Faraday rotation measurements
of distant cosmic radio sources may reveal the nature and dimension
of such primordial flux tubes if they were created in the early
universe.
\smallskip

Among all the physical processes considered for the investigation
of the energy sources of GRB, the most promising candidate appears to
be primordial flares. This can be gauged by the following 
similarities : (i) observations of solar flares and observations of GRB and,
 (ii) inferred physical parameters from  model of the primordial flare
with the observed parameters of the cosmic environment. These similarities
can also be found in the earlier work (Vahia and Rao 1988). 
\smallskip

 The GRB are similar to solar flares in the following
important observed properties : (i) the source size is compact, (ii) time scales is
$\sim$ seconds to minutes, (iii) radiation is observed in most
of the electromagnetic spectrum, (iv) non-thermal energy spectrum
and (v) polarization $< 30 \% $. 
\smallskip

In this study, we inferred from flare theory that in order to get
the required observed GRB energy, length scale of source region may be $\sim 10^{12} cms$
and strength of the magnetic field $\sim 10^{10} $ G. The difficulty
here is how to annihilate such a large scale region within few
seconds. The answer lies in the instabilities (Hood 1986) created by attaining such a 
structure of the magnetic fields. Once instability starts, reconnection
starts with in few seconds, accelerate the particles very close
to velocity of light and shock structures would  be formed which
follows the creation of non-thermal spectrum. It is also estimated
from this flare study that primordial magnetic field may be $\sim 10^{-10}$ G
which is very close to the strength of the magnetic field expected
from the observations.
It is known from solar flare observations that before eruption
of the flare, gradients of magnetic fields occur over the solar
surface. It will be interesting to get observational information regarding
development of any such strong gradients of magnetic fields (inferred from the Faraday polarization data in 
radio domain), at the site of GRB flare, before it's eruption.
In the present study we phenomenologically modeled GRB as a primordial
flare phenomenon. However, detailed solution of MHD equations is essential 
in order understand the GRB phenomenon completely.  
\smallskip

In summary, this study indicates that solar like transient MHD phenomena,
especially primordial flares in the inter galactic or inter galaxy cluster
region, may be most promising energy source for creation of 
GRB phenomena. Though most of the observed properties of GRB phenomena
and solar flare phenomena are similar, additional observational informations 
such as signature of the gradients in the cosmic magnetic fields at
the site of production of GRB is required 
in order to prove our proposed conjectures in this study.

\begin{center}
{\Large\bf Acknowledgements}
\end{center}
\noindent This paper is dedicated to my beloved parents who constantly encouraged my
research carrier when they were alive.
\begin{center}
{\Large\bf References }
\end{center}

\noindent Bagchi, J., Pislar, V. and Lima Neto, G. B., 1998, MNRAS, 296, L23

\noindent Brown, M.R., 1999, Phys. Plasma., Vol 6., No 5, 1717

\noindent Gokhale, M.H and Hiremath, K. M., 1986, Adv. Space. Res. Vol 6, No 8, pp 47-50

\noindent Haisch, B and Strong, K.T., 1991, Ann. Rev. Astron. Astrophys, 29, 275

\noindent Hood, A.W., 1986, in "Solar System Magnetic Fields", edt., E.R. Priest, 
D. Reidel Publishing Company, p. 80

\noindent Kroneberg, P.P., 1994, Rep. Prog. Phys., 57, 325

\noindent Kupurus, M., Ionson, J.A and Spicer, D.S., 1981, 19, 7-40 

\noindent Lenters, G.T and Miller, J.A., 1998, ApJ, 493, 451

\noindent Lesch, H and Chiba, M., 1997, Fundamentals of Cosmic Physics, Vol 18, No 4, 278 

\noindent Meegan, C., Hurley, A.C., Dingus, B and Matz, S., 1997, 
in "Proceedings of the IV Compton Symposium", edts., C. D. Dermer, M.S. Strickman and J.D
Kurfess, p. 407

\noindent Parker, E.N., 1994, in "Spontaneous Current Sheets in Magnetic Fields",
, Oxford Univ Press, p. 286

\noindent Petschek, H.E., 1964, "AAS-NASA Symp" in the Physics of Solar Flares,
ed. W. N. Hess, Washington, D.C., . NASA

\noindent Piran, T., 1999, Phys. Rep, in press, preprint avaialable at
http://xxx.lanl.gov/abs/astro-ph/9810256

\noindent Priest, E.R., 1981, "Solar Flare Magnetohydrodynamics", edt., E. R. Priest,
 Gordon and Breach Science Publishers, p. 14

\noindent Qu Qin-Ye and Wang Zhan-Ru, 1977, Chinese Astronomy, 1, 97

\noindent Sakurai, T., 1996, in " MHD Phenomena in Solar Atmosphere", IAU Coll. No 153, 
p. 21

\noindent Taylor, G.B and Perley, R.A., 1993, ApJ, 416, 554

\noindent Tsuneta, S and Naito, T., 1998, ApJ Let, 495, L67

\noindent Vahia, M. and Rao, A., 1988, Astron \& Astrophys, 207, 55

\noindent Vallee, J.P., MacLeaod, J.M and Broten, N.W., 1987, ApL, 25, 181

\noindent Wentzel, D.G., 1977, Sol. Phys., 52, 163-177

\end{document}